\global\def\draftcontrol{0}

   \def\versionno{ crazy modes }

\catcode`\@=11

\expandafter\ifx\csname draftcontrol\endcsname\relax\global\def\draftcontrol{0}
\fi

{\count255=\time\divide\count255 by 60
\xdef\hourmin{\number\count255}
\multiply\count255 by-60\advance\count255 by\time
\xdef\hourmin{\hourmin:\ifnum\count255<10 0\fi\the\count255}}
\def\draftdate{\number\month/\number\day/\number\year\ \ \ \hourmin }

\newcommand\makepapertitle{\par
  \begingroup
    \renewcommand\thefootnote{\@fnsymbol\c@footnote}%
    \def\@makefnmark{\rlap{\@textsuperscript{\normalfont\@thefnmark}}}%
    \long\def\@makefntext##1{\parindent 1em\noindent
            \hb@xt@1.8em{%
                \hss\@textsuperscript{\normalfont\@thefnmark}}##1}%
     \newpage
     \global\@topnum\z@   
     \@makepapertitle
     \thispagestyle{empty}\@thanks
  \endgroup
  \setcounter{footnote}{0}%
  \global\let\thanks\relax
  \global\let\makepapertitle\relax
  \global\let\@makepapertitle\relax
  \global\let\@thanks\@empty
  \global\let\@author\@empty
  \global\let\@date\@empty
  \global\let\@title\@empty
  \global\let\title\relax
  \global\let\author\relax
  \global\let\date\relax
  \global\let\and\relax
  \def\version{\let\version\@version\@gobble}
}
\def\@makepapertitle{%
  \newpage
   \ifnum\draftcontrol=1 {}
   \version\versionno
   \vskip 3em%
   \else
   \hfill\hbox to 3cm {\parbox{4cm}{\@pubnum}\hss}%
   \vskip 3em%
   \fi
   \begin{center}%
   \let \footnote \thanks
     {\LARGE {\@title}}%
     \vskip 1.5em%
     {\normalsize
       \lineskip .5em%
       \begin{tabular}[t]{c}%
         \@author
       \end{tabular}\par}%
     \vskip 1.5em%
     {\@bstract}%
     \end{center}%
     \vskip 1.5em
     \@date%
   \par
}

\gdef\@pubnum{}
\def\pubnum#1{%
  \gdef\@pubnum{#1}}

\gdef\@bstract{}
\def\Abstract#1{%
  \gdef\@bstract{%
   \parbox{\textwidth-0pc}{%
   \centerline{\bf Abstract}\penalty1000%
\kern.2cm%
\noindent
\renewcommand\baselinestretch{1.0}%
{#1}}}
}

\def\ps@paper{\let\@mkboth\@gobbletwo%
     \ifnum\draftcontrol=1
    \def\@oddfoot{\hbox to \textwidth{\tiny \versionno \hfil\tiny\draftdate}%
    \hskip -\textwidth \hbox to \textwidth{\hfil\rm\thepage\hfil}}%
     \else\def\@oddfoot{\hbox to \textwidth{\hfil\rm\thepage\hfil}}
     \fi
     \let\@evenfoot\@oddfoot
}

\def\body{\clearpage
          \pagestyle{paper}
    }

\def\@version#1{\ifnum\draftcontrol=1
\typeout{}\typeout{#1}\typeout{}
\vskip3mm\centerline{\hbox{\fbox{\normalsize{\tt DRAFT -- #1 -- }
                   {\draftdate}}}}\vskip3mm
\fi}
\let\version\@version
\long\def\eqlabel#1{\ifnum\draftcontrol=1
                    \tag@false  
                    \tag*{(\theequation) \hbox to -0.2cm{\hspace{0cm}\small{#1}\hss}}
                    \refstepcounter{equation}
                    \edef\@currentlabel{\theequation}
                    \ltx@label{#1}          
                    \else
                    \label{#1}
                    \fi
                    }
\let\st@bibitem\@bibitem
\let\st@lbibitem\@lbibitem
\ifnum\draftcontrol=1
  \def\@bibitem#1{%
    \st@bibitem{#1}\a@@label{#1}\ignorespaces}
  \def\@lbibitem[#1]#2{%
    \st@lbibitem[#1]{#2}\a@@label{#2}\ignorespaces}
  \def\a@@label#1{%
    \gdef\a@lab{\smash{\normalfont\small#1}}
    \ifvmode
      \if@inlabel
        \global\setbox\@labels\hbox{%
          \llap{\a@lab\let\a@lab\relax
                \kern\@totalleftmargin\kern\marginparsep}%
          \box\@labels}%
      \fi
    \fi}
\fi

\documentclass[12pt,letterpaper]{article}

\usepackage{amsmath,amssymb,array,calc,epsfig}

\ifnum\draftcontrol=1
\fi

\renewcommand\baselinestretch{1.25}
\renewcommand\section{\@startsection {section}{1}{\z@}%
                                   {-3.5ex \@plus -1ex \@minus -.2ex}%
                                   {2.3ex \@plus.2ex}%
                                   {\normalfont\large\bfseries}}
\renewcommand\subsection{\@startsection{subsection}{2}{\z@}%
                                   {-3.25ex\@plus -1ex \@minus -.2ex}%
                                   {1.5ex \@plus .2ex}%
                                   {\normalfont\normalsize\bfseries}}
\renewcommand\subsubsection{\@startsection{subsubsection}{3}{\z@}%
                                   {-3.25ex\@plus -1ex \@minus -.2ex}%
                                   {1.5ex \@plus .2ex}%
                                   {\normalfont\normalsize\it}}
\renewcommand\paragraph{\@startsection{paragraph}{4}{\z@}%
                                   {-3.25ex\@plus -1ex \@minus -.2ex}%
                                   {1.5ex \@plus .2ex}%
                                   {\normalfont\normalsize\bf}}

\numberwithin{equation}{section}

\def\ie{{\it i.e.}}

\def\revise#1       {\raisebox{-0em}{\rule{3pt}{1em}}%
                     \marginpar{\raisebox{.5em}{\vrule width3pt\
                     \vrule width0pt height 0pt depth0.5em
                     \hbox to 0cm{\hspace{0cm}{%
                     \parbox[t]{4em}{\raggedright\footnotesize{#1}}}\hss}}}}

\def\calj         {{\cal J}}

\def\del          {\partial}

\def\sqr#1#2{{\vcenter{\vbox{\hrule height.#2pt
 \hbox{\vrule width.#2pt height#1pt \kern#1pt
 \vrule width.#2pt}\hrule height.#2pt}}}}

\def\a{\alpha}
\def\b{\beta}
\newcommand{\qq}{\mathfrak{q}}
\newcommand{\ww}{\mathfrak{w}}

\def\e{\epsilon}

\def\dd{\delta}
\def\w{\omega}
\def\hh{\hat{h}}
\def\ha{\hat{a}}

\newcommand{\be}{\begin{equation}}

\newcommand{\ee}{\end{equation}}
\newcommand{\bea}{\begin{eqnarray}}
\newcommand{\eea}{\end{eqnarray}}

\catcode`\@=12

\begin{document}


\title{\bf The Fate of the Sound and Diffusion in Holographic Magnetic Field}
\pubnum{UWO-TH-08/17
}

\date{November 2008}

\author{
Evgeny I. Buchbinder$ ^{1}$,  Alex Buchel$ ^{1,2}$ \\[0.4cm]
\it $ ^1$Perimeter Institute for Theoretical Physics\\
\it Waterloo, Ontario N2J 2W9, Canada\\
\it $ ^2$Department of Applied Mathematics\\
\it University of Western Ontario\\
\it London, Ontario N6A 5B7, Canada
}

\Abstract{
It was shown in~\cite{BBS}
that in the presence of the magnetic field 
the sound waves in (2+1) dimensional plasma
disappear and are replaces by a 
diffusive mode. Similarly, the shear and charge diffusion fluctuations
form a subdiffusive mode. However, since the limit of small magnetic 
field does not commute with the hydrodynamic limit it is not obvious 
whether or not these modes are stable under higher order corrections.
Using AdS/CFT correspondence we show that in the case of the $M2$-brane 
plasma these modes do exist as we find the corresponding supergravity solutions. 
This allowed us to compute the conductivity and the shear viscosity 
to all orders in magnetic field. We find that the viscosity to entropy 
ratio saturates the Kovtun-Son-Starinets bound. This extends the universality 
property of the shear viscosity to the case of the strongly coupled plasma 
in external magnetic field. 
}

\makepapertitle

\body

\version\versionno


\section{Introduction}
In \cite{BBS} we studied first-order viscous magneto-hydrodynamics of strongly coupled (2+1)-dimensional conformal 
systems in the framework of gauge theory/string theory correspondence of Maldacena \cite{Juan}. 
The existence of essentially soluble holographic model, \ie, the $M2$-brane plasma \cite{Juan,GKP,Witten1},
allows one to probe intricate aspects of strongly coupled relativistic 
conformal viscous fluids in the presence of 
external magnetic field. In particular, it is known that the hydrodynamic limit 
in (2+1) dimensions does not commute 
with the limit of small magnetic field \cite{Kovtun}. As a result, one expects a drastic modification of 
the transport properties of magnetized fluids. Indeed, in \cite{BBS} is was found that the sound wave in 
magnetic plasma can propagate only in the limit of vanishing magnetic field. Depending on the scaling of the 
magnetic field in the hydrodynamic limit either only the attenuation or both the attenuation and the speed of the
sound waves are effected by the background magnetic field. 
In the latter case, it was found that the magnetic field 
reduces the effective speed of propagating sound modes, while enhancing their attenuation. 
The field theoretical arguments in the setting of Hartnoll-Kovtun-M\"uller-Sachdev (HKMS) 
magneto-hydrodynamics
\cite{Kovtun} further suggest that a hydrodynamic mode with linear 
dispersion relation disappears from the spectrum 
for finite magnetic field. Similarly, in this regime, the standard diffusive modes in viscous 
fluids become subdiffusive, 
with $\w\propto -i q^4$ dispersion. We would like to test these predictions in the holographic model 
of magneto-hydrodynamics of $M2$-brane plasma. 

Our second motivation is to understand how background fields affect the viscosity of strongly coupled 
fluids. Previously, it was discovered that relativistic holographic plasma fluids 
(with various gauge groups, matter content, with or without 
chemical potentials for conserved $U(1)$ charges, with non-commutative spatial directions) have a 
universal value of the shear viscosity at infinite 't Hooft coupling \cite{u1,Bound,u3,Alex,u5,u55} .  
The universality of the ratio of the shear viscosity to the entropy density extends also to        
non-relativistic holographic CFTs \cite{Ross, abm}. Additionally, four-dimensional conformal CFTs 
with a dual holographic description and equal central charges $a=c$ have a universal leading finite 't Hooft 
coupling correction \cite{b,bms}. On the contrary, the leading non-planar correction to the 
ratio of shear viscosity to the 
entropy density is not universal \cite{bms2}. 
Since in magnetized (2+1) fluids the shear mode becomes subdiffusive,
and the dual holographic setting falls outside the most general universality class considered in \cite{Alex},
one naturally questions whether the ratio of shear viscosity to the entropy density continues to be universal.

The paper is organized as follows. 
In the next section, we review magneto-hy\-dro\-dy\-na\-mics of HKMS and 
its soluble holographic realization as hydrodynamics of dyonic black holes in M-theory.  
In section 3, we extend the supergravity analysis 
of \cite{BBS} and discuss propagation of 'sound waves' in $M2$-brane plasma in the 
hydrodynamic limit with finite 
external magnetic field. In section 4, we study 'shear modes' and compute the ratio 
of the shear viscosity to the entropy density 
of the $M2$-brane plasma in the external magnetic field. In section 5, we comment on
a computation of the shear viscosity using the 
Kubo formula. Some technical details are presented in Appendix A. 

After this work was completed, the paper~\cite{Hansen} appeared 
which approaches magneto-hydrodynamics from the gravity side along the lines 
of~\cite{Minwalla}. In the future,  
it would be interesting to compare the results of~\cite{Hansen} with the HKMS approach. 


\section{Magneto-Hydrodynamics and Dyonic Black Hole Geometry}


\subsection{Hydrodynamic Modes in the Presence of Magnetic Field}


In this section, we will review magneto-hydrodynamics in 
(2+1) dimensions following~\cite{BBS}. 
We are interested in hydrodynamic properties of the (2+1) 
dimensional theory on the large number of $M2$-branes
in the presence of the external magnetic field. This theory can be 
understood as the maximally supersymmetric gauge theory in three dimensions 
at the infrared fixed point. The equations of motion and the conformal properties
can be rigorously derived using the fact that the theory admits a holographic dual 
description as M-theory on $AdS_4 \times S^7$~\cite{Juan, GKP, Witten1}. 
The appropriate field theory equations then follow from the symmetries 
of the $AdS$ background. This was studied in detail in~\cite{BBS}
and here we will quote the results. The relevant field theory equations of
motion are just the conservation laws of the form
\begin{equation}
\begin{split}
&\partial^{\nu}T_{\mu \nu} =F_{\mu \nu}J^{\nu}\,, \\
&\partial_{\mu} J^{\mu}=0\,,
\end{split}
\eqlabel{1.1}
\end{equation}
where $T_{\mu \nu}$ is the stress-energy tensor, $J^{\mu}$ is the current and $F^{\mu \nu}$ 
is the {\it external} electromagnetic field. 
In the present paper, it is taken to be magnetic, that is 
\begin{equation}
F_{0 i}=0\,, \qquad i=1, 2\,, \qquad F_{i j}=\epsilon_{i j}B\,.
\eqlabel{1.1.1}
\end{equation}
Another important equation is 
\begin{equation}
T^{\mu}_{\ \mu}=0\,,
\eqlabel{1.2}
\end{equation}
which means that the field theory under study is conformal. 
Eq.~\eqref{1.2}
is the consequence of the fact that 
the leading near-the-boundary asymptotics
of the gauge field in $AdS$ is constant, which implies that the magnetic 
field represents a marginal deformation. See~\cite{BBS} for details. 

The expressions for $T^{\mu \nu}$ and $J^{\nu}$ to first order 
in derivatives were derived in~\cite{Kovtun} from 
postulating positivity of the entropy production along the lines 
of Landau and Lifshitz~\cite{LL}. The stress-energy tensor 
is given by the standard expression 
\begin{equation}
T^{\mu \nu} =\e u^{\mu} u^{\nu} +P\Delta^{\mu \nu} -\eta
(\Delta^{\mu \a}\Delta^{\nu \b} (\del_{\a}u_{\b}+\del_{\b}u_{\a})-\Delta^{\mu \nu}
\del_{\gamma}u^{\gamma})-\zeta\Delta^{\mu \nu}\del_{\a} u^{\a}\,.
\eqlabel{1.3}
\end{equation}
Here 
\begin{equation}
\Delta^{\mu \nu}=\eta^{\mu \nu}+u^{\mu}u^{\nu}\,,
\eqlabel{1.3.1}
\end{equation}
$u^{\mu}$ is the fluid $3$-velocity, $\e$ and $P$ are the energy density and the pressure 
respectively, and $\eta$ and $\zeta$ are the shear and bulk viscosity. Since our theory is 
conformal it follows that $\zeta=0$. It is important to note that $P$ is different 
from the thermodynamic pressure $p$~\cite{Kovtun}, 
\begin{equation}
P=p- M B\,,
\eqlabel{1.3.2}
\end{equation}
where $M$ is the magnetization. Also note that conformal invariance 
implies that 
\begin{equation}
c_s^2=\frac{\del P}{\del \e}=\frac{1}{2}\,.
\eqlabel{1.4}
\end{equation}
Similarly, the current $J^{\mu}$ is given by 
\begin{equation}
J^{\mu}=\rho u^{\mu} +\sigma_Q\Delta^{\mu \nu}(-\del_{\nu}\mu+F_{\nu \a}u^{\a} 
+\frac{\mu}{T} \del_{\nu}T)\,,
\eqlabel{1.5}
\end{equation}
where $\rho$ is the charge density, $\mu$ is the chemical potential, $T$ is the temperature
and $\sigma_Q$ is the conductivity coefficient. To study fluctuations around the equilibrium 
state 
\begin{equation}
u^{\mu}=(1, 0, 0)\,, \quad T={\rm const.}\,, \quad  \mu={\rm const.}\,,
\eqlabel{1.6}
\end{equation}
we choose $(\dd u_1=\dd u_x$, $\dd u_2=\dd u_y$, $\dd T$, $\dd \mu)$ as the independent 
quantities. As usual, all the fluctuations are of the plane-wave form 
$exp(-i \omega t+i q y)$. In this paper, we are interested in hydrodynamics
with no net charge density and, correspondingly, with no chemical potential
\begin{equation}
\rho=0\,, \qquad \mu=0\,. 
\eqlabel{1.7}
\end{equation}
In this case, as was shown in~\cite{BBS}, the equations for the linear fluctuations
get separated into the two decoupled pairs. The first pair reads
\begin{equation}
\begin{split}
& \w \left(\frac{\del \e}{\del T}\right)_{\mu} \delta T- q (\e +P)\delta u_y=0\,, \\
& \w (\e+P) \delta u_y -q \left(\frac{\del P}{\del T}\right)_{\mu}\delta T
+i q^2 \eta \delta u_y+ i \sigma_Q B^2 \delta u_y=0\,.
\end{split}
\eqlabel{1.8}
\end{equation}
If we set $B=0$ these equations describe the sound waves with dispersion 
relation 
\begin{equation}
\w=\pm \frac{q}{2} -i q^2\frac{\eta}{\e+P}\,,
\eqlabel{1.9}
\end{equation}
where eq.~\eqref{1.4} has been used. 
We will refer to these equations as to the ``sound channel''.
The second decoupled pair of equations is 
\begin{equation}
\begin{split}
& \w (\e+P)\delta u_x -q B \sigma_Q \delta \mu +
i \sigma_Q B^2 \delta u_x +i q^2 \eta \delta u_x=0 \,,\\
& \w \left( \frac{\del \rho}{\del \mu}\right)_{T} \delta \mu +
q \sigma_Q B \delta u_x +i q^2 \sigma_Q \delta \mu=0\,.
\end{split}
\eqlabel{1.10}
\end{equation}
Note that here we are assuming that the susceptibility  
$ \left( \frac{\del \rho}{\del \mu}\right)_{T}$ does not vanish 
at $\rho=\mu=0$. 
If we set $B=0$ these two equations further decouple. One equation describes
the shear mode $\delta u_x$ with dispersion relation
\begin{equation}
\w=-i q^2 \frac{\eta}{\e +P}\,.
\eqlabel{1.11}
\end{equation}
The other one describes the charge diffusion mode $\delta \mu$ with dispersion 
relation
\begin{equation}
\w=-i q^2 \frac{\sigma_Q}{\left(\frac{\del \rho}{\del \mu}\right)_{T}}\,.
\eqlabel{1.12}
\end{equation}
We will refer to eqs.~\eqref{1.10} as to the ``shear channel''. 
If  we turn on the magnetic field $B$, the hydrodynamic 
modes undergo a drastic change. The reason, as one can see from
eqs.~\eqref{1.8} and~\eqref{1.10}, is that the limit of small $B$ does not commute 
with the hydrodynamic limit of small $\w$ and $q$. So the magnetic field cannot be thought of as
a small perturbation. In the case of non-zero $B$ 
(kept fixed in the hydrodynamic limit)
we obtain the following solutions. 
In the sound channel we do not get the sound waves anymore. Instead, we obtain a constant 
solution 
\begin{equation}
\w =-i \sigma_Q \frac{B^2}{\e+P}\,,
\eqlabel{1.13}
\end{equation}
and a diffusive mode
\begin{equation}
\w =-\frac{i q^2}{2} \frac{\e+P}{\sigma_Q B^2}\,.
\eqlabel{1.14}
\end{equation}
One can interpret~\eqref{1.14} as that the effective speed of sound vanishes once 
the magnetic field is turned on. In the shear channel, the usual shear and diffusive 
modes disappear. Instead, we also obtain a constant 
solution~\eqref{1.13} and a subdiffusive mode
\begin{equation}
\w=- i q^4 \frac{\eta}{B^2\left(\frac{\del \rho}{\del \mu}\right)_{T}}\,.
\eqlabel{1.15}
\end{equation}
The modes~\eqref{1.14} and~\eqref{1.15} will be the main focus of our paper. 

Since the limit of small $B$ does not commute with the hydrodynamic
limit one can worry that the solutions~\eqref{1.14} and~\eqref{1.15} cannot be trusted.
Indeed, these solutions imply a hierarchy of amplitudes. From eqs.~\eqref{1.8} and~\eqref{1.14} 
it follows that 
\begin{equation}
\frac{\delta T}{\delta u_y} \sim  \frac{1}{q}\,,
\eqlabel{1.16}
\end{equation}
and from eqs.~\eqref{1.10} and~\eqref{1.15} it follows that
\begin{equation}
\frac{\delta \mu}{\delta u_x} \sim  \frac{1}{q}\,. 
\eqlabel{1.17}
\end{equation}
Hence, given the amplitudes of the linearized fluctuations $\delta u_x$ 
and $\delta u_y$, we find that the amplitudes of the linearized
fluctuations $\delta T$ and $\delta \mu$ are strongly enhanced in the
hydrodynamic limit. 
Then one can expect that the solutions~\eqref{1.14} 
and~\eqref{1.15} are unstable under higher order (higher derivative) corrections
to $T^{\mu\nu}$ and $J^\mu$.
That is, terms which are naively 
of higher order because they are suppressed by higher powers of $\w$ and $q$ 
can, in fact, modify hydrodynamic equations at lower order because they come with a large 
amplitude. 

Unfortunately, at the level of the effective field theory it is very difficult to answer
whether or not these solutions exist. However, for the case of the $M2$-brane plasma we can use
the description in terms UV complete M-theory 
on $AdS_4 \times S^7$ background. Our gravitational
analysis in the later sections will show that in the case of the $M2$-brane 
plasma with large number of $M2$-branes
the modes~\eqref{1.14} and~\eqref{1.15} do exist. 
Moreover, finding these solutions on the gravity side will allow us 
to calculate the conductivity coefficient $\sigma_Q$ and the shear 
viscosity $\eta$ to all orders in magnetic field.
These results indicate that
AdS/CFT correspondence is a helpful method to study hydrodynamic modes 
whose very existence is subtle from the field theory prospective. 


\subsection{Supergravity Magneto-Hydrodynamics}


According to AdS/CFT correspondence, in the limit when the number 
of $M2$-branes becomes very large, their dynamics can be described 
by the eleven-dimensional supergravity on $AdS_4 \times S^7$. 
For our purposes, this theory theory can be consistently truncated 
to Einstein-Maxwell theory on $AdS_4$~\cite{Son}. The supergravity action 
is then given by\footnote{For simplicity, we set the radius of $AdS_4$ to unity.}
\begin{equation}
S=\frac{1}{g^2}\int d^4 x \sqrt{-g} \left[ -\frac{1}{4}R +\frac{1}{4}F_{M N}F^{M N}-\frac{3}{2}
\right]\,,
\eqlabel{1.18}
\end{equation}
where the bulk coupling constant $g$ is given by 
\begin{equation}
\frac{1}{g^2}=\frac{\sqrt{2}N^{3/2}}{6 \pi}\,,
\eqlabel{1.19}
\end{equation}
where $N$ is the number of $M2$-branes. The corresponding equations of motion are 
\begin{equation}
\begin{split}
&R_{M N}=2 F_{M L}F_N^L-\frac{1}{2} g_{M N} F_{L P}F^{L P} -3 g_{M N}\,,\\
&\nabla_{M}F^{M N}=0\,.
\end{split}
\eqlabel{1.20}
\end{equation}
The equilibrium state of magneto-hydrodynamics is described by 
(asymptotically $AdS_4$) dyonic 
black hole geometry with planar horizon whose Hawking temperature is identified 
with the plasma temperature. The solution looks as follows~\cite{HK}
\begin{equation}
\begin{split}
&d s^2 =-c_1(r)^2 d t^2 +c_2(r)^2(d x^2 +d y^2)+c_3(r)^2 d r^2\,,\\
& F= h \a^2 dx \wedge d y +q \a d r \wedge d t\,,
\end{split}
\eqlabel{1.21}
\end{equation}
where 
\begin{equation}
c_1(r)^2=\frac{\a^2}{r^2}f(r)\,, \quad c_2(r)^2=\frac{\a^2}{r^2}\,,  
\quad c_3(r)^2=\frac{\a^2}{f(r) r^2}\,,
\eqlabel{1.22}
\end{equation}
and 
\begin{equation}
f(r)=1+ (h^2+q^2) r^4 -(1+h^2 +q^2) r^3\,.
\eqlabel{1.23}
\end{equation}
In these coordinates, $r=1$ corresponds to the horizon and $r=0$ is the boundary. 
The black hole parameters $(h, q, \a)$ are related to the field theory 
magnetic field, chemical potential and temperature as~\cite{HK}
\begin{equation}
B=h \a^2\,, \quad \mu =-q \a\,, \quad T=\frac{\a}{4 \pi}(3-h^2-q^2)\,.
\eqlabel{1.24}
\end{equation}
Note that the $x y$-component of $F$ goes to a constant $h \a^2$ 
on the boundary and is identified with the boundary theory magnetic field $B$. 
On the other hand, the $t$-component of the vector potential behaves near the 
boundary as $A_t =-q \a r$. It is interpreted as the chemical potential
in the boundary theory. 

Now we list some thermodynamic properties of the dyonic black hole.  
See~\cite{HK} for more details. The appropriate thermodynamic potential is obtained
by evaluating the (renormalized) action~\eqref{1.18} and is given by 
\begin{equation}
\Omega=-p V =V \frac{1}{g^2} \frac{\a^3}{4}\left(-1-
\frac{\mu^2}{\a^2}+3 \frac{B^2}{\a^4}\right)\,,
\eqlabel{1.25}
\end{equation}
where $V$ is the area of the $(x, y)$-plane and $p$ is the thermodynamic pressure. 
The other quantities of importance are the density of energy, entropy and 
electric charge. They are 
given by 
\begin{equation}
\e=\frac{1}{g^2} \frac{\a^3}{2}\left( 1+ \frac{\mu^2}{\a^2}+\frac{B^2}{\a^4}\right)\,,
\eqlabel{1.26}
\end{equation}
\begin{equation}
s=\frac{\pi}{g^2} \a^2\,,
\eqlabel{1.27}
\end{equation}
and
\begin{equation}
\rho=\frac{1}{g^2} \a \mu\,.
\eqlabel{1.28}
\end{equation}
In addition, we introduce magnetization per unite area
\begin{equation}
M=-\frac{1}{V} \left( \frac{\del \Omega}{\del B}\right)_{T, \mu}=-\frac{1}{g^2}\frac{B}{\a}\,. 
\eqlabel{1.29}
\end{equation}
Just like on the field theory side, we introduce
\begin{equation}
P=p-M B\,.
\eqlabel{1.30}
\end{equation}
One can show~\cite{HK} that it is $P$ rather than $p$ that coincides 
with the spatial components $\langle T^{x x}\rangle$ and $\langle T^{y y}\rangle$
of the stress-energy tensor, just like we have in eq.~\eqref{1.3}. It is straightforward
to check that 
\begin{equation}
P=\frac{\e}{2}\,,
\eqlabel{1.31}
\end{equation}
which is consistent with conformal invariance. 
In this paper, we consider magneto-hydrodynamics in the absence of the net charge density $\rho$. 
Thus, we set $q=0$. Note that even though $\rho$ and $\mu$ vanish 
the derivative
\begin{equation}
\left(\frac{\del \rho}{\del \mu}\right)_T= \frac{\a}{g^2} \,.
\eqlabel{1.32}
\end{equation}
is non-zero. 

To study a holographic dual of the hydrodynamic modes, we
need to find linear fluctuations of the supergravity equations 
of motion~\eqref{1.20} around the black hole background~\eqref{1.21},
\begin{equation}
\begin{split}
& g_{M N} \to g_{M N} + h_{M N}\,, \\
& A_{M} \to A_{M} + a_{M} \,.
\end{split}
\eqlabel{1.33}
\end{equation}
It is convenient to impose the gauge 
\begin{equation}
h_{t r}=h_{x r}=h_{y r}=h_{r r}=0\,, \quad a_r=0\,.
\eqlabel{1.33.1}
\end{equation}
In parallel with field theory, the fluctuations $g_{M N}$ and $a_{M}$ 
will be of the form $exp(-i \w t+i q y)$ and the $r$-dependence is to be 
obtained from solving the linearized Einstein and Maxwell equations~\eqref{1.20}. 
As was explained in~\cite{BBS}, for both $q$ and $h$ non-zero, 
all the metric and gauge field fluctuations couple to each other
and no decoupling of various modes exist. The reason is that the background~\eqref{1.21}
does not have any symmetry which usually allows one to decouple scalar-, vector- 
and tensor-type fluctuations. However, in the case of interest $q=0$ there exist two sets 
of decoupled fluctuations. The first set corresponds to the field theory sound channel. 
The corresponding fluctuations are 
\begin{equation}
\{h_{t t}, h_{t y}, h_{x x}, h_{y y}, a_x\}\,.
\eqlabel{1.34}
\end{equation}
The second set corresponds to the field theory shear channel
and includes the following fluctuations
\begin{equation}
\{h_{t x}, h_{x y}, a_{t}, a_y\}\,.
\eqlabel{1.35}
\end{equation}
In the next section, we will show that the fluctuations~\eqref{1.34}
indeed correctly describe the diffusive mode~\eqref{1.14}. 
In section 4, we will show that the fluctuations~\eqref{1.35} indeed describe 
the subdiffusive mode~\eqref{1.15}. 


\section{The Fate of the Sound Waves}




In this section, we will consider the equations of motion for the
fluctuations~\eqref{1.34}. Let us introduce 
\begin{equation}
\begin{split}
h_{tt}=&c_1(r)^2\ \hh_{tt}=e^{-i\w t+iq y}\ c_1(r)^2\  H_{tt}\,,\\
h_{ty}=&c_2(r)^2\ \hh_{ty}=e^{-i\w t+iq y}\ c_2(r)^2\  H_{ty}\,,\\
h_{xx}=&c_2(r)^2\ \hh_{xx}=e^{-i\w t+iq y}\ c_2(r)^2\  H_{xx}\,,\\
h_{yy}=&c_2(r)^2\ \hh_{yy}=e^{-i\w t+iq y}\ c_2(r)^2\  H_{yy}\,,\\
a_x=&i e^{-i\w t+iq y}\ \ha_x\,,
\end{split}
\eqlabel{2.1}
\end{equation} 
where $H_{tt}$, $H_{ty}$, $H_{xx}$, $H_{yy}$ and, $\ha_x$ are functions of the radial coordinate only
and $c_1(r)$ and $c_2(r)$ are defined in eqs.~\eqref{1.22} and~\eqref{1.23}.
Expanding eqs.~\eqref{1.20} to linear order we obtain the following system
of equations
\begin{equation}
\begin{split}
0=&H_{tt}''+H_{tt}'\ \left[\ln\frac{c_1^2c_2}{c_3}\right]'+
\frac 12\left[H_{xx}+H_{yy}\right]'\ \left[\ln \frac{c_2}{c_1}\right]'
-\frac{c_3^2}{2c_1^2}\biggl(q^2\frac{c_1^2}{c_2^2}\ (H_{tt}+H_{xx})\\
&+\w^2\ (H_{xx}+H_{yy})+2\w q\ H_{ty}\biggr)
-3\ \frac{c_3^2}{c_2^4}\ h^2\a^4\ (H_{xx}+H_{yy})+6\ \frac{c_3^2}{c_2^4}\ h\a^2 q\ \ha_x\,,
\end{split}
\eqlabel{2.2}
\end{equation}
\begin{equation}
\begin{split}
0=&H_{ty}''+H_{ty}'\ \left[\ln\frac{c_2^4}{c_1c_3}\right]'
+\frac{c_3^2}{c_2^2}\ \w q\ H_{xx}-4 \frac{c_3^2}{c_2^4}\ h\a^2 \left(h\a^2\ H_{ty}+\w\ \ha_x\right)\,,
\end{split}
\eqlabel{2.3}
\end{equation}
\begin{equation}
\begin{split}
0=&H_{xx}''+\frac 12 H_{xx}'\ \left[\ln\frac{c_1^5c_2}{c_3^2}\right]'+
\frac 12\  H_{yy}'\ \left[\ln \frac {c_2}{c_1}\right]'
+\frac{c_3^2}{2c_1^2}\biggl(\w^2(H_{xx}-H_{yy})-q^2\frac{c_1^2}{c_2^2}( H_{tt}+H_{xx})\\
&-2\w q H_{ty}\biggr)
-\frac{c_3^2}{c_2^4}\ h^2\a^4\ (H_{xx}+H_{yy})+2\ \frac{c_3^2}{c_2^4}\ h\a^2 q\ \ha_x\,,
\end{split}
\eqlabel{2.4}
\end{equation}
\begin{equation}
\begin{split}
0=&H_{yy}''+\frac 12 H_{yy}'\ \left[\ln\frac{c_1^5c_2}{c_3^2}\right]'+
\frac 12\  H_{xx}'\ \left[\ln \frac {c_2}{c_1}\right]'
+\frac{c_3^2}{2c_1^2}\biggl(\w^2(H_{yy}-H_{xx})+q^2\frac{c_1^2}{c_2^2}( H_{tt}-H_{xx})\\
&+2\w q H_{ty}\biggr)
-\frac{c_3^2}{c_2^4}\ h^2\a^4\ (H_{xx}+H_{yy})+2\ \frac{c_3^2}{c_2^4}\ h\a^2 q\ \ha_x\,,
\end{split}
\eqlabel{2.5}
\end{equation}
\begin{equation}
\begin{split}
0=&\ha_x''+\ha_x'\ \left[\ln\frac {c_1}{c_3}\right]'+
\frac{c_3^2}{c_1^2}\ \ha_x\left(\w^2-\frac{c_1^2}{c_2^2} q^2\right)
+ \frac{c_3^2}{2c_2^2}\ h\a^2\ \biggl(q(H_{tt}+H_{xx}+H_{yy})\\
&+2 \w\ \frac{c_2^2}{c_1^2}\ H_{ty}\biggr)\,. 
\end{split}
\eqlabel{2.6}
\end{equation}
In addition, we obtain three first class constraints from 
varying the action with respect to the gauge fixed metric components 
$h_{t r}$, $h_{y r}$ and $h_{r r}$
\begin{equation}
\begin{split}
0=&\w\left(\left[H_{xx}+H_{yy}\right]'+
\left[\ln\frac{c_2}{c_1}\right]'\ (H_{xx}+H_{yy})\right)+
q\left(H_{ty}'+2\left[\ln\frac{c_2}{c_1}\right]'\ H_{ty}\right)\,,
\end{split}
\eqlabel{2.7}
\end{equation}
\begin{equation}
\begin{split}
0=&q\left(\left[H_{tt}-H_{xx}\right]'-
\left[\ln\frac{c_2}{c_1}\right]'\ H_{tt}\right)+\frac{c_2^2}{c_1^2}\w\ H_{ty}'
+4\ h\a^2\ \frac{\ha_x'}{c_2^2}\,,
\end{split}
\eqlabel{2.8}
\end{equation}
\begin{equation}
\begin{split}
0=&[\ln c_1c_2]'\left[H_{xx}+H_{yy}\right]'-[\ln{c_2^2}]'\ H_{tt}'+\frac{c_3^2}{c_1^2}
\biggl(\w^2\ (H_{xx}+H_{yy})+2\w q\ H_{ty}\\
&+q^2\ \frac{c_1^2}{c_2^2}\left(H_{tt}-H_{xx}\right)\biggr)
-2\frac{c_3^2}{c_2^4}\ h^2\a^4\ (H_{xx}+H_{yy})+4 \frac{c_3^2}{c_2^4}\ h\a^2q\ \ha_x\,.
\end{split}
\eqlabel{2.9}
\end{equation}
If we set in these equations $h=0$, we see that we can also consistently set $\ha_x=0$.
Then the remaining equations for  $H_{tt}$, $H_{ty}$, $H_{xx}$ and $H_{yy}$
can be shown to coincide with those in~\cite{Herzog2} and describe the sound waves 
with dispersion relation~\eqref{1.9}. See~\cite{Herzog2} for details. 
To continue, we note that the gauge~\eqref{1.33.1} does not fully fix
the diffeomorphism invariance. Since we have five (second order in derivatives) equations and 
three (first order in derivatives) constraints, there are precisely two 
combinations invariant under the residual gauge transformations. They were found in~\cite{BBS} to be 
\begin{equation}
\begin{split}
Z_H=&4\frac{q}{\w} \ H_{ty}+2\ H_{yy}-
2 H_{xx}\left(1-\frac{q^2}{\w^2}\frac{c_1'c_1}{c_2'c_2}\right)+2\frac{q^2}{\w^2}
\frac{c_1^2}{c_2^2}\ H_{tt}\,,\\
Z_A=&\ha_x+\frac{1}{2q}\ h\a^2\ \left(H_{xx}-H_{yy}\right)\,.
\end{split}
\eqlabel{2.10}
\end{equation}
Then from eqs.~\eqref{2.2}-\eqref{2.6} and~\eqref{2.7}-\eqref{2.9} 
we obtain two decoupled gauge invariant equations for $Z_H$ and $Z_A$
\begin{equation}
\begin{split}
0=&A_HZ_H''+B_HZ_H'+C_HZ_H+D_HZ_A'+E_HZ_A\,, 
\end{split}
\eqlabel{2.11}
\end{equation}
\begin{equation}
\begin{split}
0=&A_AZ_A''+B_AZ_A'+C_AZ_A+D_AZ_H'+E_A Z_H\,. 
\end{split}
\eqlabel{2.12}
\end{equation}
The connection coefficients $\{A_H,\cdots,E_{A}\}$ 
can we computed from eqs.~\eqref{2.2}-\eqref{2.9}
and \eqref{2.10} using explicit expressions for the $c_i$'s in~\eqref{1.22}. 
Since these coefficients are very long and cumbersome
we will not present them in the paper.
Below we will present these equations in the limit of small $\omega$ and $q$. 

As the next step, we will discuss the boundary conditions. According to the 
general prescription~\cite{Andrei1, Andrei2}, in order to obtain 
the dispersion relation (poles in the retarded Green's function) we have to impose 
the following boundary conditions
\begin{itemize}
\item 
$Z_H$ and $Z_A$ are incoming waves at the horizon $r=1$. 
\item 
$Z_H$ and $Z_A$ satisfy the the Dirichlet boundary conditions on the 
boundary $r=0$. That is, both $Z_H$ and $Z_A$ have to vanish at $r=0$. 
\end{itemize}
In~\cite{BBS} it was shown that $Z_H$ and $Z_A$ have the following behavior 
at the horizon
\begin{equation}
\begin{split}
& Z_H (r)=f(r)^{-i \ww/2}z_H(r)\,,\\
& Z_A (r)=f(r)^{-i \ww/2}z_A(r)\,,
\end{split}
\eqlabel{2.13}
\end{equation}
where we introduce 
\begin{equation}
\ww=\frac{\w}{2 \pi T}\,, \quad \qq=\frac{q}{2 \pi T}\,.
\eqlabel{2.14}
\end{equation}
The functions $z_H$ and $z_A$ are now regular and non-vanishing at the horizon. 
In addition, they have to satisfy the Dirichlet boundary conditions at $r=0$. 

A crucial ingredient in search for the solution is a proper understanding 
of the correct relative normalization. Since our equations~\eqref{2.11} and~\eqref{2.12}
are homogeneous, we can normalize one of the functions, say $z_H$, to be unity at the horizon $r=1$. 
However, after that we cannot normalize $z_A$. Moreover, the ratio
$\frac{z_A}{z_H}$ can depend on $q$. Since we are going to solve the equations 
of motion perturbatively in $q$ it is important to establish how $\frac{z_A}{z_H}$
scales with $q$. We recall that we are looking for a diffusive solution with 
$\omega \sim q^2$. Then from eq.~\eqref{2.10} it follows that for small $q$ 
\begin{equation}
z_H \sim \frac{1}{q^2}H_{t t}\,, \quad z_A \sim \ha_x\,.
\eqlabel{2.16}
\end{equation}
Note that due to an additional symmetry between $x$ and $y$
at $q=0$ it follows that $H_{x x}=H_{y y}$ at $q=0$
and the leading term for small $q$ in $z_A$ is $\ha_x$. Furthermore, we know that 
$H_{t t}$ is dual to the $t t$-component of the boundary stress-energy tensor
whereas $\ha_x$ is dual to the $x$-component of the boundary current $J^x$. Going back to the 
field theory side, using eqs.~\eqref{1.3} and~\eqref{1.5} one can show that 
\begin{equation}
\frac{\delta J^x}{\delta T^{t t}}\sim \frac{\delta u_y}{\delta T}\sim q \,,
\eqlabel{2.17}
\end{equation}
where eq.~\eqref{1.16} has been used. 
Then we find that $\frac{z_A}{z_H} \sim q^3$. Let us now parameterize 
the ansatz for our solution. We parameterize the dispersion relation as follows
\begin{equation}
\ww =-i \frac{\qq^2}{h^2} C\,.
\eqlabel{2.18}
\end{equation}
The additional $h^2$ in the denominator is dictated by the 
field theory result~\eqref{1.14}. The coefficient $C$ is now assumed to have a
perturbative expansion in $h$. Similarly, it is convenient to pull out the appropriate 
powers in $h$ in the $\qq$-expansion of $Z_H$ and $Z_A$. We end up with the following ansatz 
\begin{equation}
\begin{split}
& Z_H=f(r)^{-i \ww/2}\left( F_1(r)+\frac{\qq^2}{h^2} F_3(r)+ {\cal O}(\qq^4)\right)\,, \\
& Z_A=f(r)^{-i \ww/2} \frac{\qq^3}{h^5}\left( F_2(r)+ {\cal O}(\qq^2)\right)\,.
\end{split}
\eqlabel{2.19}
\end{equation}
One can show that with these powers of $h$ in the denominators, the functions
$F_i(r)$ have now a perturbative expansion in $h^2$. 
Note that since $z_H$ is normalized to be unity at the horizon, we can choose 
$F_1(r)=1$ and $F_3(r)=0$ at $r=1$. 

First, we will solve eqs.~\eqref{2.11} and~\eqref{2.12} to leading order in $h$, 
that is ignoring the $h$-dependence in $C$ and $F_i(r)$. 
Doing this we will determine that $C$ can be fixed entirely by imposing
the proper boundary condition on the fluctuations $Z_H$ and $Z_A$ near the
horizon. Thus, we will be able to generalize the procedure and find an
analytic expression for $C$ to all orders in $h$ without actually
finding the full analytic solution for the fluctuations.
We start with the equations to the leading order in $\qq$. From eq.~\eqref{2.11} 
we obtain
\begin{equation}
F_1^{\prime \prime}-\frac{2}{r} F_1^{\prime}=0\,.
\eqlabel{2.20}
\end{equation}
The solution with the prescribed above boundary conditions is 
\begin{equation}
F_1(r)=r^3\,.
\eqlabel{2.21}
\end{equation}
From eq.~\eqref{2.22} we get 
\begin{equation}
F_2^{\prime \prime} -\frac{2+r^3}{r(1-r^3)}F_2^{\prime} +
\frac{3 \alpha C^2}{8 r (1-r^3)}=0\,,
\eqlabel{2.22}
\end{equation}
where the solution for $F_1$~\eqref{2.21} have been used.
The general solution to this equation is given by 
\begin{equation}
\begin{split}
&F_2(r)= C_1+ C_2 \ln (1-r^3)  \\
&+\frac{\a C^2}{32} \left[
2 \sqrt{3}  \arctan \left(\frac{1 +2 r}{\sqrt{3}}\right) 
- 2  \ln(1-r) +  \ln(1+r+r^2)\right]\,.
\end{split}
\eqlabel{2.23}
\end{equation}
The integration constant $C_2$ has to be fixed by requiring that $F_2(r)$ is regular at the horizon. 
It gives
\begin{equation}
C_2=\frac{\a C^2}{16}\,.
\eqlabel{2.24}
\end{equation}
The other integration constant $C_1$ is fixed by requiring that $F_2(r)$ vanishes at $r=0$
to be 
\begin{equation}
C_1=- \frac{\a \pi C^2}{32 \sqrt{3}}\,.
\eqlabel{2.25}
\end{equation}
Note that we are not able to fix the diffusion constant $C$ working at leading 
order in $\qq$. We have to go to next-to-leading order in $\qq$ and consider 
the equation for $F_3$\footnote{One can worry that going to next-to-leading order
in $\qq$ cannot fix $C$ because the equation for $F_3$ will also depend on higher 
order coefficients in the dispersion relation. However, it is straightforward to check 
that it is not the case.}
\begin{equation}
F_3^{\prime \prime} -\frac{2}{r} F_3^{\prime} +
\frac{r^4(81 (-4+r^3)-432 C(-2+r^3) +512 C^2 r(-1+r^3))}{96 (1-r^3)^2} =0\,,
\eqlabel{2.26}
\end{equation}
where the above solutions for $F_1(r)$ and $F_2(r)$ have been used. 
It is possible to find the general solution for $F_3(r)$. We will not write it here 
because it is rather lengthy. 
The solution has a logarithmic singularity at the horizon of the form
\begin{equation}
F_3(r)\sim \frac{1}{32} (-9 +16 C) \ln (1-r)\,.
\eqlabel{2.27}
\end{equation}
Requiring that the solution is smooth fixes $C$ to be 
\begin{equation}
C= \frac{9}{16}+ {\cal O}(h)\,.
\eqlabel{2.28}
\end{equation}
The two integration constants in the solution for $F_3(r)$ are fixed by requiring that 
$F_3(r)$ vanishes at $r=0$ and $r=1$. 

From this procedure it becomes clear that to obtain $C$ we need to understand the near horizon 
structure of the solution and require that it is smooth. 
It is possible to perform this analysis for arbitrary $h$. The details are presented in Appendix A.
This allows us to find the exact value of $C$
\begin{equation}
C=\frac{3}{16} (3-h^2)(1+h^2)\,. 
\eqlabel{2.29}
\end{equation}
Thus, we have managed to reproduce the diffusive mode in the sound channel on the gravity side!
We proved that this mode does exist and the exact (to all orders in $h$) value of the diffusion 
constant is given by eq.~\eqref{2.29}. 

To summarize the results, we have obtained a supergravity solution with the 
following dispersion relation
\begin{equation}
\ww =-\frac{i \qq^2}{h^2} \frac{3}{16}(3-h^2)(1+h^2) +{\cal O}(\qq^4)\,. 
\eqlabel{2.30}
\end{equation}
Let us compare it with the field theory counterpart~\eqref{1.14}. 
For this we will rewrite~\eqref{1.14} in terms of $(\ww, \qq, \a, h)$. 
From eqs.~\eqref{2.14}, \eqref{1.24} and \eqref{1.26} it follows that~\eqref{1.14}
can be written as
\begin{equation}
\ww =-\frac{i \qq^2}{h^2} \frac{3}{16} (3-h^2)(1+h^2) \frac{1}{g^2 \sigma_Q}\,. 
\eqlabel{2.31}
\end{equation}
Comparing~\eqref{2.31} and~\eqref{2.30} we obtain the following answer 
for the conductivity coefficient
\begin{equation}
\sigma_Q=\frac{1}{g^2}= \frac{\sqrt{2}N^{3/2}}{6 \pi}\,.
\eqlabel{2.32}
\end{equation}
This coincides with the result for $\sigma_Q$ obtained earlier 
in~\cite{Kovtun, HH} to {\it leading order} in $B$. Our result~\eqref{2.32}, however, 
is valid to {\it all orders} in $B$. 
Our calculation provides a rigorous proof that the conductivity 
coefficient does not depend on the magnetic field. 


\section{The Fate of the Shear Modes and Charge Diffusion}


In this section, we will consider the fluctuations~\eqref{1.35} describing 
the shear channel.
Let us introduce 
\begin{equation}
\begin{split}
h_{t x}=& e^{-i\w t+iq y}\ c_2(r)^2\  H_{t x}\,,\\
h_{x y}=& e^{-i\w t+iq y}\ c_2(r)^2\  H_{x y}\,,\\
a_{t}=& i e^{-i\w t+iq y} \ha_{t}\,,\\
a_y=&i e^{-i\w t+iq y}\ \ha_y\,,
\end{split}
\eqlabel{3.1}
\end{equation} 
where $H_{t x}, H_{x y},  \ha_{t}$ and $\ha_{y}$ are functions of the radial coordinate $r$. 
Expanding eqs.~\eqref{1.20} to linear order we obtain the following system of equations 
\begin{equation}
\begin{split}
0= & H_{t x}^{\prime \prime} + \left[\ln \frac{c_2^4}{c_1 c_3}\right]^{\prime}  H_{t x}^{\prime}
-\frac{c_3^2}{c_2^4} (4 h^2 \alpha^4 + q^2 c_2^2)H_{t x}-\frac{q \omega c_3^2}{c_2^2} H_{x y} \\
- & \frac{4 h \a^2 q c_3^2}{c_2^4}\ha_{t} -\frac{4 h \a^2 \omega c_3^2}{c_2^4}\ha_{y}\,, 
\end{split}
\eqlabel{3.2}
\end{equation}
\begin{equation}
0= H_{x y}^{\prime \prime} +\left[\ln \frac{c_1 c_2^2}{c_3}\right]^{\prime}  H_{x y}^{\prime}
+\frac{\omega^2 c_3^2}{c_1^2} H_{x y} +  \frac{q \w c_3^2}{c_1^2}H_{t x}\,,
\eqlabel{3.3}
\end{equation}
\begin{equation}
0= \ha_{t}^{\prime \prime} +\left[\ln \frac{c_2^2}{c_1 c_3}\right]^{\prime} \ha_{t}^{\prime}
-\frac{q^2 c_3^2}{c_2^2}\ha_t -\frac{q \w c_3^2}{c_2^2}\ha_y -\frac{h \a^2 q c_3^2}{c_2^2}H_{t x}\,,
\eqlabel{3.4}
\end{equation}
\begin{equation}
0= \ha_{y}^{\prime \prime} +\left[\ln \frac{c_1}{c_3}\right]^{\prime} \ha_{y}^{\prime}
+\frac{\w^2 c_3^2}{c_1^2}\ha_y +\frac{q \w c_3^2}{c_1^2}\ha_t +\frac{h \a^2 \w c_3^2}{c_1^2}H_{t x}\,.
\eqlabel{3.5}
\end{equation}
In addition, we have two first class constraints obtained from varying the action with respect 
to the gauge fixed components $h_{x r}$ and $a_r$
\begin{equation}
0=\frac{q}{2} H_{x y}^{\prime} +\frac{\w c_2^2}{c_1^2}H_{t x}^{\prime} +\frac{2 h \a^2}{c_2^2}\ha_{y}^{\prime}\,,
\eqlabel{3.6}
\end{equation}
and 
\begin{equation}
0=\frac{q}{c_2^2} \ha_{y}^{\prime} +\frac{\w}{c_1^2}\ha_{t }^{\prime} \,.
\eqlabel{3.7}
\end{equation}
If we set $h=0$ we see that the equations describing $(H_{tx}, H_{xy})$ and $(\ha_t, \ha_y)$ 
decouple from each other. In this case, the equations for $(H_{tx}, H_{xy})$ describe 
the shear modes with dispersion relation~\eqref{1.11}~\cite{Herzog1}. Similarly, 
the equations for $(\ha_t, \ha_y)$ describe diffusion with dispersion relation~\eqref{1.12}~\cite{Herzog1}. 

To continue, we introduce quasinormal modes invariant under the residual diffeomorphisms
\begin{equation}
\begin{split}
& Z_H= q H_{t x}+ \omega H_{x y} \,, \\
& Z_A= q \ha_t +\w \ha_y -\frac{\w}{q}h \a^2 H_{x y}\,.
\end{split}
\eqlabel{3.8}
\end{equation}
Then from eqs.~\eqref{3.2}-\eqref{3.8} we obtain two decoupled gauge invariant 
second order equations for $Z_H$ and $Z_A$ of the form~\eqref{2.11}, \eqref{2.12}. 
These equations are rather lengthy and we will not write them in the paper. 
Below, we will present them to the lowest orders in $\omega$ and $q$. 
The quasinormal modes $Z_H$ and $Z_A$
have the same boundary conditions as discussed in the previous section. 
Namely, $Z_H$ and $Z_A$ are incoming waves at the horizon $r=1$
and vanish on the boundary $r=0$. Repeating the same analysis as in the previous section, 
we arrive at the following ansatz for our solution
\begin{equation}
\begin{split}
& Z_H (r)= f(r)^{-i \ww/2}\frac{\qq}{h} \left( F_1(r)+ \qq^2 F_3(r)+ \frac{\qq^4}{h^2}F_5(r)+ 
{\cal O}(\qq^6) \right)\,, \\
& Z_A (r)= f(r)^{-i \ww/2}  \left( F_2(r)+ \qq^2 F_4(r)+ \frac{\qq^4}{h^2}F_6(r)+ 
{\cal O}(\qq^6) \right)\,,
\end{split}
\eqlabel{3.9}
\end{equation}
where 
\begin{equation}
\ww=\frac{\w}{2 \pi T}\,, \quad \qq=\frac{q}{2 \pi T}\,,
\eqlabel{3.10}
\end{equation}
and the functions $F_i(r)$ are non-singular and non-vanishing at the horizon 
and have a perturbative expansion in $h$. In addition, they satisfy the Dirichlet
conditions on the boundary. Since the equations are homogeneous, we can choose 
the following normalization 
at the horizon $r=1$
\begin{equation}
F_2(r)|_{r=1}=1\,, \quad F_4(r)|_{r=1} =F_6(r)|_{r=1}=\ldots =0\,.
\eqlabel{3.10.1}
\end{equation}
Once~\eqref{3.10.1} is chosen, no further normalization condition for $F_1$, $F_3$, $F_5, \ldots$ at $r=1$ 
can be imposed. We parameterize the dispersion relation as follows
\begin{equation}
\ww=-i \frac{\qq^4}{h^2} C\,,
\eqlabel{3.11}
\end{equation}
where the diffusion constant $C$ is assumed to have a perturbative expansion in $h$. 
As in the previous section, we will first solve the equations 
to leading order in $h$, that is 
ignoring the $h$ 
dependence in $C$ and $F_i(r)$. Then we will generalize our method for arbitrary $h$ 
and find $C$ to all orders in $h$. 
To leading order in $\qq$ we obtain the following simple equations 
\begin{equation}
F_1^{\prime \prime} -\frac{2}{r} F_1^{\prime}=0\,,
\eqlabel{3.12}
\end{equation}
and 
\be 
F_2^{\prime \prime}=0\,. 
\eqlabel{3.13}
\ee
The solution with the prescribed above boundary conditions is 
\begin{equation}
F_1 (r) = b_1 r^3\,, \quad F_2(r)=r\,.
\eqlabel{3.14}
\end{equation}
The integration constant $b_1$ so far is not fixed. Our analysis shows that 
it is fixed by requiring that the function $F_5(r)$ is non-singular at the horizon. 
We will not present the details of this analysis since they are not important for our purposes. 
At the next order in $\qq$ we obtain equations for $F_3$ and $F_4$. They can be
solved analytically 
but these solutions are not of interest for us. It turns out that the diffusion constant 
$C$ is determined from the equation for $F_6$ which reads 
\begin{equation}
F_6^{\prime \prime} +\frac{16 C^2 r^2 }{9 (1-r^3)}F_2^{\prime \prime}
-\frac{C r( 27 r (-1+r^3) +16 C (2+r^3))}{9 (-1+r^3)^2} F_2^{\prime}
+\frac{3 C r (2+r^2)}{2 (-1+r^3)^2}F_2=0\,.
\eqlabel{3.15}
\end{equation}
Note that to determine $F_6$ we need to know only $F_2$ which was found in~\eqref{3.14}. 
The other functions $F_1$, $F_3$, $F_4$ do not enter this equation.
It possible to find the general solution for $F_6$ analytically. We will not write
it because it is rather lengthy. It turns out that to determine $C$ it is enough 
to study the behavior of $F_6$ near the horizon. We find that near the horizon 
$F_6$ has a logarithmic singularity of the form
\begin{equation}
F_6(r)=\frac{C}{54}(27-32 C)\ln (1-r) +\ldots\,,
\eqlabel{3.16}
\end{equation}
where the ellipsis stands for the non-singular terms. Since $F_6$ has to be smooth 
at the horizon it follows that we can have a non-trivial solution
\be
C=\frac{27}{32}+ {\cal O}(h)\,.
\eqlabel{3.17}
\ee
The two integration constants of $F_6$ are fixed by requiring that $F_6$ vanishes 
at $r=0$ and $r=1$. 

Now it is clear how to generalize the above procedure for arbitrary $h$. 
The diffusion constant $C$ is obtained from requiring that the solution
does not have a logarithmic singularity at the horizon. In fact, it possible to carry out 
the near-horizon analysis for arbitrary $h$. 
Repeating similar steps as in Appendix A, we obtain 
the final result 
\begin{equation}
C=\frac{(3-h^2)^3}{32}\,.
\eqlabel{3.18}
\end{equation}
Thus, we have managed to reproduce the subdiffusive mode~\eqref{1.15} in the
shear channel on the gravity side! We proved that it does exist and found the
value of the diffusive constant $C$ to {\it all orders} in 
magnetic field in the large $N$ limit. 

To summarize, we have obtained the supergravity solution with the following dispersion 
relation
\be
\ww=-i \frac{\qq^4}{h^2} \frac{(3-h^2)^3}{32}\,.
\eqlabel{3.19}
\ee
Let us compare it with the field theory counterpart~\eqref{1.15}. 
Using eqs.~\eqref{3.10}, \eqref{1.24} and~\eqref{1.32} we can rewrite~\eqref{1.15}
in the form 
\begin{equation}
\ww=-i \frac{\qq^4}{h^2} \frac{(3-h^2)^3}{8 \a^2} \eta g^2\,.
\eqlabel{3.20}
\ee
The factor $(3-h^2)^3$ comes from rewriting the temperature in terms of $\alpha$ and $h$. 
Comparing~\eqref{3.19} and~\eqref{3.20} we conclude that 
\be
\eta=\frac{\a^2}{4 g^2}\,, 
\eqlabel{3.21}
\ee
where, to recall, $\alpha$ is related to the temperature and the magnetic field as follows
\be
T= \frac{\a}{4 \pi} \left( 3- \frac{B^2}{\a^4}\right)\,.
\eqlabel{3.22}
\ee
Note that despite the simple form~\eqref{3.21}, $\eta$, after being rewritten in terms of $T$ and $B$,  
has a non-trivial dependence on the magnetic field. Recalling the expression 
for the entropy density~\eqref{1.27} we, finally, obtain
\be
\frac{\eta}{s}=\frac{1}{4 \pi}\,.
\eqlabel{3.23}
\end{equation}
We see that the ratio $\eta/s$ saturates the KSS bound~\cite{Bound}. This extends 
the universality theorem of~\cite{Alex} for the case of strongly coupled plasma in 
external magnetic field. 


\section{The Kubo Formula}


In the previous section, we extracted the ratio of the shear viscosity to the entropy ratio 
from the dispersion relation of the shear quasinormal modes. 
Alternatively, one can use a Kubo formula, as in 
\cite{Alex}, to directly evaluate the shear viscosity:
\begin{equation}
\eta=\lim_{\w\to 0}\frac {1}{2\w i}\left[G_{xy,xy}^{A}(\w,0)-G_{xy,xy}^R(\w,0)\right] \,,
\eqlabel{ek}
\end{equation}
where $G_{xy,xy}^{A}$ ($G_{xy,xy}^{R}$) is the advanced (retarded) two-point correlation function of the 
stress-energy  tensor (with indicated spatial indices) evaluated 
at zero momentum. Naively, the universality 
arguments presented in \cite{Alex} do not hold here. Indeed, 
the dual gravitational mode, $h_{xy}$ does not generically decouple 
as one can see from eqs. \eqref{3.2}-\eqref{3.5}.
Furthermore, the background of the bulk vector field (holographically dual to the background boundary 
magnetic field) is not solely polarized along the time direction, as assumed in \cite{Alex}. 
In this section, we revisit the argument of \cite{Alex} and extend the universality theorem 
to the case of background magnetic field in (2+1)-dimensional strongly coupled plasma. 

First, notice that even though the gravitational mode $h_{xy}$ does not decouple for non-zero momentum 
$q$, see eqs.~\eqref{3.2}-\eqref{3.5}, the decoupling occurs for $q=0$, 
which is all what is needed for the computation of the 
correlation functions in \eqref{ek}. Physically, the reason why such a decoupling 
occurs is because for vanishing $q$
there is an additional symmetry in plasma associated with a reflection  along the $y$-axis. 
As a result, the 
graviton polarization $h_{xy}$ is the only fluctuating mode which is doubly parity odd under reflections along the
$x$- and $y$-axis. Hence, it must decouple. Second, from \eqref{3.3} 
we see that the rescaled (see eq. \eqref{3.1}) graviton 
wavefunction $H_{xy}$ at $q=0$ satisfied the equation of motion for the minimally coupled massless scalar 
in a regular Schwarzschild horizon geometry \eqref{1.21}. 
But now, we are precisely in the setup of the 
universality arguments of \cite{Alex}! One can literally repeat the analysis 
presented there to establish that 
the shear viscosity $\eta$, as determined by \eqref{ek}, is proportional to the 
entropy density $s$ with 
the proportionality coefficient as in \eqref{3.23}.


\section*{Acknowledgments}

Research at Perimeter Institute is supported by the
Government of Canada through Industry Canada and by the Province of
Ontario through the Ministry of Research \& Innovation. A.B.
gratefully acknowledges further support by an NSERC Discovery grant
and support through the Early Researcher Award program by the
Province of Ontario. 


\appendix
\section{Exact value of $C$}
For a background magnetic field $h$ held fixed in the hydrodynamic limit, 
the sound channel dispersion 
relation is given by \eqref{2.18}, with the constant $C\equiv C(h)$. 
Here, we outline steps necessary to obtain exact analytical expression for $C(h)$. 

Within the ansatz \eqref{2.19}, $F_1$ is given by \eqref{2.21}. From \eqref{2.11} 
and \eqref{2.12} 
we further find the following equations for $F_2$ and $F_3$
\begin{equation}
\begin{split}
0=F_2''+\frac{2 r^4 h^2-r^3 (h^2+1)-2}{r (r^3 h^2-r^2-r-1) (r-1)}F_2'-
\frac{(h^2-3) \alpha C^2}{8r (r^3 h^2-r^2-r-1) (r-1)}\,,
\end{split}
\eqlabel{f2eq}
\end{equation} 
\begin{equation}
\begin{split}
0=&F_3''-6 \frac{r^4 h^2-1}{r (r^4 h^2-3)} F_3'+\frac{12 r^2 h^2}{r^4 h^2-3} F_3
\\
&-\frac{r^2 (h^2-3) (r^3 (h^2+1)-4) (r^4 h^4+r^4 h^2-16 h^2 r+9 h^2+9)}{
2C^2 \alpha (r^3 h^2-r^2-r-1) (r-1) (r^4 h^2-3)} F_2'\\
&-\frac{24 r^2 (h^2-3) h^2}{C^2 \alpha (r^4 h^2-3)} F_2+\calj_3\,,
\end{split}
\eqlabel{f3eq}
\end{equation} 
where 
\begin{equation}
\begin{split}
\calj_3=&\frac{r^2}{32(r-1)^2 (r^3 h^2-r^2-r-1)^2 (r^4 h^2-3)}\biggl(
512 r^3 (r-1) (r^3 h^2-r^2-r-1) C^2\\
&+16 r^2 (-54 (h^2+1)+84 h^2 r+27 (h^2+1)^2 r^3-60 h^2 (h^2+1) r^4+24 h^4 r^5\\
&+3 h^2 (h^2+1)^2 r^7-6 h^4 (h^2+1) r^8+4 r^9 h^6) C-(h^2-3)^2 (-36 (h^2+1)+40 h^2 r\\
&+9 (h^2+1)^2 r^3-32 h^2 (h^2+1) r^4+48 h^4 r^5+h^2 (h^2+1)^2 r^7-12 h^4 (h^2+1) r^8\\
&+8 r^9 h^6)
\biggr)\,.
\end{split}
\eqlabel{j3}
\end{equation}
It is straightforward to construct power series solutions first for $F_2$ \eqref{f2eq} and then for $F_3$
\eqref{f3eq} near the horizon.  Regularity of $\{F_2,F_3\}$ 
for small $x=1-r$ then uniquely fixes $C$ as in \eqref{2.29}.



\end{document}